# A Simple Iterative Approach for Constant Chemical Potential Simulations at Interfaces

Ademola Soyemi, Khagendra Baral, and Tibor Szilvási[*]

Department of Chemical and Biological Engineering, The University of Alabama, Tuscaloosa, Alabama 35487, United States

* Email: tibor.szilvasi@ua.edu

Abstract

Chemical potential of species in solution is essential for understanding various chemical processes at interfaces. Molecular dynamics (MD) simulations, constrained by fixed compositions, cannot satisfy a constant chemical potential condition as solute species can migrate to the interface and deplete the bulk due to solute-interface interactions. In this study, we introduce a simple and computationally efficient approach named iterative constant chemical potential molecular dynamics (iCμMD) simulation, which helps simulate targeted molar concentrations of species in solution. iCμMD overcomes the limitations of conventional MD by adjusting the number of species in the solution to reach a target concentration (chemical potential). We demonstrate our approach using solid-liquid and liquid-air interfacial systems as case studies. Specifically, we perform classical force field-based MD simulations of $NaCl_{aq}$-air and $NaCl_{aq}$-graphite interfaces and machine learning interatomic potential (MLIP)-based MD simulations of the $Na_2SO_{4,aq}$-graphene interface. Our results show that the iCμMD approach efficiently achieves the desired bulk ion concentration within two iterations and can also be integrated with MLIP-driven simulations which enable constant potential simulations with DFT-level accuracy. We show that iCμMD offers a robust and simple computational framework for constant chemical potential simulations as its only requirement is to be able to converge interfacial simulations with a measurable bulk region.

1. Introduction

Chemical potential is a fundamental thermodynamic property that drives key phenomena like reaction kinetics, phase transitions, and diffusion of ions, thus influencing a system's macroscopic behavior.[1] Chemical potential is crucial in catalysis, electrochemistry, materials science, and biological systems, where chemical potential governs fundamental phenomena at interfaces.[2] While driving forces such as coulombic interactions or adsorption lead to accumulation of species at the interface, the chemical potential remains constant across the whole system (i.e., the chemical potential at the interface equals that in the bulk region) as the interface and bulk region are in equilibrium. Moreover, the bulk region serves as a reservoir and any species present should exchange without depleting the reservoir.

Computational models generally fail to simulate constant chemical potential due to the finite size of the simulations. To overcome this limitation, the grand canonical Monte Carlo (GCMC)[3-4] and hybrid GCMC/molecular dynamics (MD)[5-7] methods are widely employed for simulating open systems where particles are exchanged between a simulation cell and an external reservoir. GCMC/MD methods offer key advantages, such as enabling direct control over particle number and chemical potential, making them suitable for studying systems with varying particle counts or concentration. However, the GCMC/MD methods become highly inefficient in dense fluids, where the probability of successful particle insertion becomes very small,[5] for example at solid-liquid interfaces. As such, modeling solid-liquid interfaces at constant chemical potential remains a significant challenge using GCMC/MD methods.

Recently, Finney et al.[8-9] introduced an approach known as the constant chemical potential molecular dynamics (CµMD) method to study solid-liquid interfaces which mimics the open-boundary systems.[8-9] In their CµMD setup, the simulation cell is divided into four contiguous regions, namely the interfacial region, the control region where the solution concentration is monitored and held fixed, the reservoir region which provides a large reservoir of species to adjust the concentration of species in the control region, and the force region (0.2 nm) which lies between the control and reservoir regions and enables the addition of species from the reservoir region to the control region. When species are depleted from the control region due to migration of species to the interfacial region, species are released to the control region by applying a force to species in the force region thereby maintaining a fixed concentration in the control region. Using CµMD, Finney and coworkers[10-11] simulated graphite-electrolyte systems at constant chemical potential of ions using classical force fields.

Classical force fields are commonly used to model solid-liquid interfaces due to their ability to reach the necessary time scales and system sizes to obtain converged thermodynamic properties.[8-9, 11-12] However, classical force fields can suffer from inaccuracy due to the lack of parameters for the complex interactions between the solid and liquid or solid and solutes. *Ab initio* molecular dynamics (AIMD) studies using density functional theory (DFT) would be ideal due to high accuracy, however, AIMD studies of solid-liquid interfaces are limited due to the short time and length scales currently accessible to DFT-based AIMD. To circumvent the limitations of AIMD and classical force fields, machine learning interatomic potentials (MLIPs) that are trained on DFT reference data have recently gained popularity since they can in principle be used to carry out DFT-quality simulations while gaining access to longer time and length scales.[13-22] As such, DFT-quality simulations of solid-liquid interfaces are becoming increasingly tractable with the advent of MLIPs, and thus it also becomes critical to integrate MLIPs with simulation protocols that allow for maintaining constant chemical potential.

In this work, we introduce the iterative CµMD (iCµMD) simulation approach. The iCµMD approach does not require a reservoir region or complicated simulation setup and it can be applied for classical or MLIP-driven MD simulations. Figure 1 outlines the general approach, showing that the targeted molar concentration is reached by adjusting the numbers of species through successive iterations. In each new iteration, multiple species are added to or removed from the system based on the concentration and number of species in the bulk region. This procedure is then repeated until the targeted bulk concentration is reached. We show results can be obtained equivalent to CµMD simulation within just two iterations by applying iCµMD. We demonstrate our method using the same graphite-electrolyte ($NaCl_{aq}$) system from Finney et al.[8-9] and an electrolyte ($NaCl_{aq}$)-air interfaces modeled with a classical force field with a target bulk concentration of 1 M. Additionally, we consider a more complex graphene-electrolyte ($Na_2SO_{4,aq}$) system modeled using a MLIP. iCµMD offers several advantages, including ease of setup, computational efficiency, and compatibility with MLIPs.

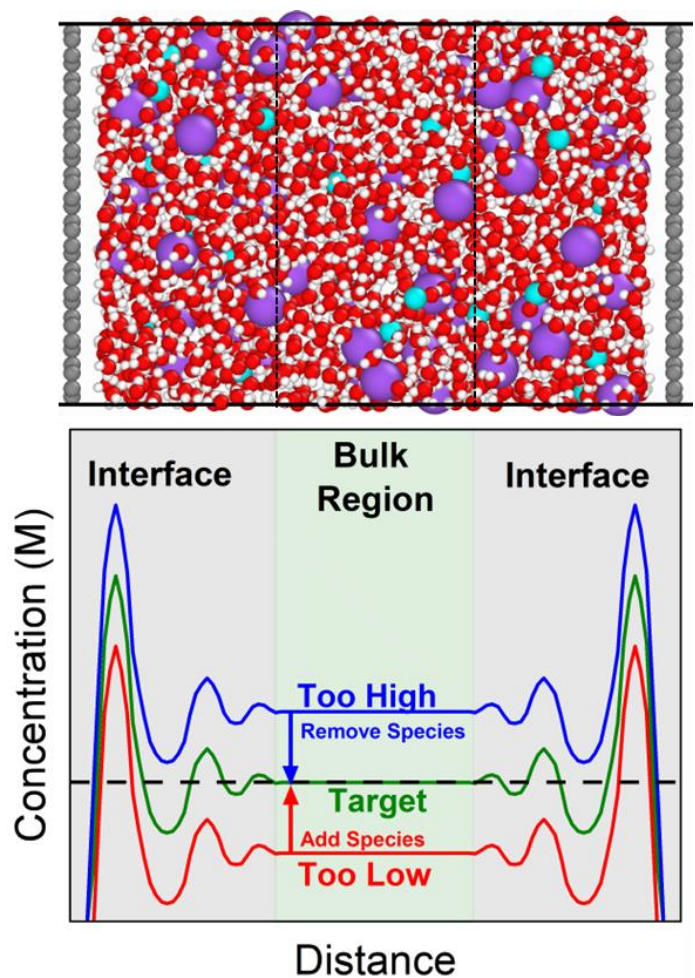

**Figure 1**. Schematic representation of the iCµMD approach. The top panel shows a snapshot of the $Na_2SO_{4,aq}$ - graphene system, while the bottom panel outlines the iCµMD approach in general. The vertical dashed black lines and horizontal solid black lines in the top panel represent the bulk region boundaries, and simulation cell boundaries, respectively. Atom color code: C = Grey, Na = Purple, S = Cyan, O = Red and H = White.

2. The iCµMD Approach

MD in the NVT ensemble fixes the number of species, therefore the bulk region concentration will be dictated by the interactions between the species and the surface which can cause segregation of the species to the solid-liquid interface and their depletion in the bulk. In many applications, such as electrochemistry, the chemical potential of ions remains constant in the bulk region of the system even as reactions occur at, and ions saturate the solid-liquid interface. Thus, to overcome the limitations of GCMC, hybrid MC/MC methods and CµMD, we propose the iCµMD approach. The iCµMD approach offers a simple yet practical alternative whereby rather than sampling insertions and deletions on the fly via GCMC, we periodically measure the bulk concentration from a relatively short MD run then iteratively adjust the total count by removing or inserting species. By iterating this 'run-measure-update' loop until the bulk concentration converges to the desired target concentration, iCµMD can achieve a constant-µ condition in a few iterations and with minimal changes to the conventional MD workflow. The remainder of this section describes how we define the bulk region, the bulk concentration convergence criteria that determines when to stop an iteration, the update rule for changing the total number of species and the iCµMD stopping criteria. We focus on describing the general iCµMD method in this section, while more specific details on the MD runs and analysis of ion distributions for the studied systems are given in the Computational Methods section.

2.1. Definition of the Bulk Region and Measurement of Bulk Concentration

A critical first step in iCµMD is to define the 'bulk' region of the simulation cell. In practice, we exclude the interfacial region, typically within 10 Å of the confining surface, and compute the bulk concentration within a region around the center of the solution, for example ±5 Å from the center of the solution (See Computational Methods for specific definition of the bulk region used for the NaCl$_{aq}$- graphite, NaCl$_{aq}$ - air, and Na$_2$SO$_{4,aq}$ - graphene systems studied in this work). Defining interfacial and bulk region can be done in general by calculating the concentration of species as a function of the distance from the surface and checking for a region of constant concentration.

2.2. Bulk Concentration Convergence Criteria and Ion Number Update

Based on the definition of the bulk region given above, we measure the bulk concentration of species at regular intervals. To assess whether the bulk concentration has truly plateaued, we compute the slope of the moving average of the bulk concentration and deem an iteration converged when the slope becomes zero with fluctuations within a user-defined threshold. Additionally, we require that the slope remains within the slope threshold for a specific amount time (defined by the user) to avoid transient convergence. We emphasize that different systems (e.g. viscous solutions, or solid-gas interfacial systems) may exhibit different bulk concentration convergence or fluctuations and as such parameters such as the moving average window length and slope tolerance and the overall definition of convergence are flexible, and future users can define the convergence criteria to suit their needs and specific application (See Computational Methods for specific convergence criteria used for the NaCl$_{aq}$- graphite, NaCl$_{aq}$ - air, and Na$_2$SO$_{4,aq}$ - graphene systems studied in this work).

When an iteration step is considered converged and the bulk concentration is higher or lower than the target bulk concentration (within threshold), we use the following equation to update the number of species:

$$N_t^{i+1} = \frac{C_b^t}{C_b^i} N_t^i \qquad (1)$$

where, $N_t^i$ is the total number of target species on the i$^{th}$ iteration, $N_t^{i+1}$ is the total number of target species on the (i+1)$^{th}$ iteration, $C_b^t$ is targeted molar concentration of species in the bulk of the solution, and $C_b^i$ is resulting bulk concentration of the target species on the i$^{th}$ iteration. Overall, if the bulk concentration is higher than the target bulk concentration, $N_t^{i+1} - N_t^i$ $(or\ \Delta N)$ number of species are removed from the solution and this new simulation setup is used to initialize the next iteration. Meanwhile, if the bulk concentration is lower than the target bulk concentration, we add more species to initialize the next iteration by replacing $\Delta N$ solvent molecules with $\Delta N$ species to be added. If $\Delta N$ is not an integer, we round to the nearest whole number. Finally, to define the stopping criteria for the iCµMD simulations, we stop iterating when the bulk concentration is within a given user defined threshold generally related to the fluctuation of the converged bulk concentration in a given iteration.

3. Computational Methods

In this section, we discuss the specific definitions and simulations setups related to the NaCl$_{aq}$- graphite, NaCl$_{aq}$ - air, and Na$_2$SO$_{4,aq}$ – graphene systems used as demonstrations in this work.

3.1. Classical Force Field MD

To illustrate our iCµMD simulation approach, we construct the NaCl$_{aq}$-graphite and NaCl$_{aq}$-air interfaces as representative models and set our target concentration to 1 M. The graphite structure was adopted from Finney's work,[8, 23] comprising eight stacked layers of graphene centered along the z-axis of the simulation cell. The initial solution was then created using Packmol[24] with a nominal 1 M molar concentration of ions (See Figure 2) in a simulation cell of dimensions 53.93 Å x 55.17 Å x 148.07 Å. The initial simulation cell contained a total of 43,759 atoms comprising of 8 stacked graphene layers comprising of 1,144 carbons each (9,152 carbon atoms in total), 11,389 water molecules (34,167 atoms in total), and 220 NaCl ion pairs (440 atoms in total). We also applied a uniform charge of +0.01 and -0.01 e/atom to the carbon atoms in the outermost graphite layers to drive the formation of an electrical double layer. For the NaCl$_{aq}$-air interface, we generate the initial configuration using Packmol and center the solution within the simulation cell (See Figure 2) of dimensions 53.93 Å x 55.17 Å x 148.07 Å. The initial simulation cell contained a total of 34,607 atoms comprising of 11,389 water molecules (34,167 atoms in total), and 220 NaCl ion pairs (440 atoms in total).

MD simulations were performed employing the GROMACS code (version 2024.1)[25] and force field parameters as in Finney et al.[10, 26] within the NVT ensemble at 298 K using a time step of 2 fs for the ionic motion. The system temperature was controlled by the velocity rescaling thermostat with a time constant of 0.1 ps. The Lennard–Jones potential and electrostatic interactions were calculated with a cutoff distance of 0.9 nm, and the particle mesh Ewald (PME) method was used for long-range electrostatic interaction with a Fourier spacing of 0.1 nm. Cross-term interactions were approximated with the Lorentz–Berthelot mixing rules. Water molecules were represented by the SPC/E model [27] with an O−H bond length of 1 Å and an H−O−H angle of 109.47°. Additionally, the O-H bonds were constrained using the LINCS algorithm[28] in GROMACS. Na$^+$ and Cl$^−$ ions were modeled using Joung and Cheatham force fields.[29] The C−C bonds and C−C−C angles were fixed to 1.42 Å and 120° in the graphite layer, respectively. The graphite-water interactions were modeled using the atom pair potential from Wu and Aluru,[30] and graphite-ions (Na$^+$ and Cl$^-$) interactions were modeled using the atom pair potential from Williams et al.[31] The graphite layers were held rigid during the MD simulations. Initial simulations were started with a nominal ion concentration of 1 M as a reference in graphite-electrolyte and electrolyte-air interfaces. Following an initial relaxation via the steepest-descent algorithm, we performed a 10 ns equilibration before production simulations for the first iteration were run for a duration of 50 ns. It should be noted that for the graphite-electrolyte, equilibration simulations were run without the applied charges on the outermost graphite layers. We analyze the Z-direction distribution of ions every 1 ns and define the bulk region as 5 Å from the edge of both ends of the simulation cell (i.e., the center of the solution) in the Z-direction (i.e., a 10 Å region) for the graphite-NaCl$_{aq}$ system. Similarly, for the NaCl$_{aq}$-air system, we define the bulk region as ±5 Å from the center of the solution (i.e., a 10 Å region). It should the noted that the ion density in the Z-direction is taken in 1 Å bins. To define the iteration stopping criteria, we use a moving average window of 5 ns, and a slope threshold of ±2 x 10$^{-3}$ M/ns for the NaCl$_{aq}$-graphite and NaCl$_{aq}$-air systems. Additionally, we require that the slope remains within the threshold for a duration of at least 15 ns to avoid transient convergence. Finally, to define the stopping criteria for the iCµMD simulations, we stop iterating when the bulk concentration is within ±0.03 M of the target concentration based on the uncertainty defined previously by Finney et al.[10]

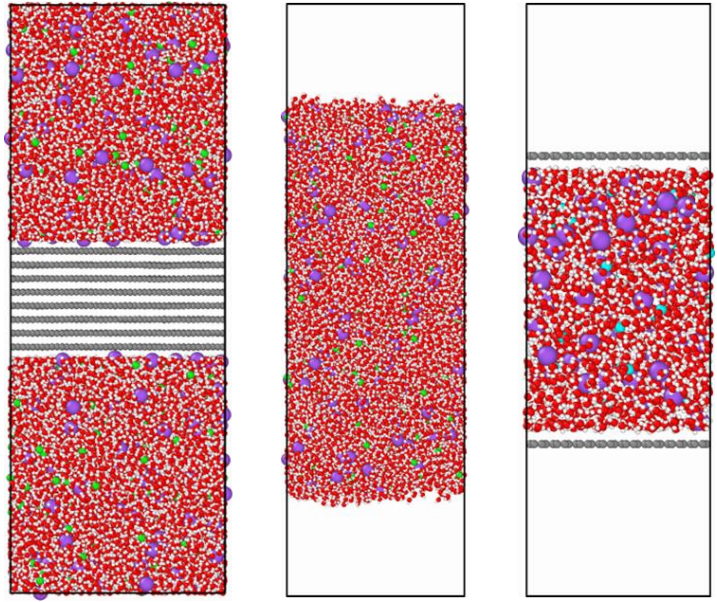

**Figure 2**. Snapshots of $NaCl_{aq}$- graphite (left), $NaCl_{aq}$ - air (middle), and $Na_2SO_{4,aq}$ - graphene (right) systems. The simulation cell for the $NaCl_{aq}$- graphite and $NaCl_{aq}$ - air systems have dimensions of 53.93 Å x 55.17 Å x 148.07 Å, respectively. The $NaCl_{aq}$- graphite system was initialized with a total of 43,759 atoms, while the $NaCl_{aq}$ - air system was initialized with a total of 34,607 atoms. The simulation cell for the $Na_2SO_{4,aq}$ - graphene system has dimensions of 34.15 Å x 29.57 Å x 110.00 Å and was initialized with a total of 5,785 atoms. See Computational methods for per-specie counts. Atom color code: C = Grey, Na = Purple, Cl = Green, S = Cyan, O = Red, and H = White.

3.2. MLIP Training Methodology

All reference DFT calculations were performed using the Vienna Ab initio Simulation Package (VASP) version 6.3.2.[32-33] The RPBE exchange- correlation functional[34] was used, and the electron-core interaction was represented by projector augmented waves (PAWs) with a plane wave basis energy cutoff of 520 eV. In all DFT single-point calculations, the Brillouin zone was sampled using a gamma k-point grid and the with electronic occupations smeared using a Gaussian smearing width of 0.03 eV. To obtain accurate forces, the energy convergence criteria was set to $10^{-7}$ eV. Building on previous studies[35-36] that demonstrated dispersion-corrected RPBE accurately describes bulk water we include dispersion corrections using Grimme's D3 method with zero damping.[37] In addition to this, a recent study has shown that the D3 method inaccurately calculates the $C_{AB}^6$ term for pairs involving cations such as $Na^+$.[38] Consequently, we utilized a modified D3(0) whereby the contributions of $Na^+$ to the dispersion correction are removed.

The accuracy of an MLIP in reproducing a correct potential energy surface (PES) is greatly determined by the training set.[19-20, 39] Since the PES does not have a physically derived functional form, the accuracy of the MLIP is therefore limited to configurations it is supplied in the training set. Thus, in order to learn a correct PES, all relevant parts of the configuration space must be well sampled from when curating a training set. Specifically, all subsystems such as bulk water, sodium sulfate in bulk water (at different concentrations initialized using Packmol[24]), and slab systems such as water on graphene, sodium sulfate in water on graphene (at different concentrations) must be well represented in the training set. To accomplish this, we carried out 0.5 ns NVT MD simulations at different densities for bulk systems using the MACE-MP-0 pretrained MLIP[40] via the Atomic Simulation Environment (ASE, version 3.22.1).[41] These simulations were run with deuterated hydrogens at 300 K under Langevin dynamics and a 1 fs timestep. All MD simulations of bulk water were run using 128 water molecules and 0 to 5 $Na_2SO_4$ ions to simulate a concentration range of 0 to 2 M. The densities of the simulation cells were also varied with in ±4 % of experimental densities.[42] Additionally, since NVT simulations typically provide equilibrium configurations we carried out constant potential energy MD simulations (also known as contour exploration)[43] whereby we explicitly generate high energy structures which have been shown to improve the accuracy and stability of the resulting MLIP.[19] Here, we perform constant potential energy MD simulations, using the bulk water systems (with and without ions), with a max step size of 1 Å for 10,000 steps at each energy level up to 5 eV above the energy of the optimized geometry with an energy step size of 0.5 eV. Single point energy calculations were performed on uncorrelated structures chosen at 3 ps intervals and were then used as the initial training and validation set. Similarly, for the slab models we ran 0.5 ns NVT MD with a sodium sulfate concentration range of 0 – 2 M. All MD simulations were run with a graphene sheet containing 96 carbon atoms of dimension of 17.075 x 14.787 Å and 128 water molecules. Additionally, 20 Å of vacuum was included to minimize the effect of dipole interactions between periodic images.

We train MLIPs using NequIP package.[44] The development branch of NequIP downloaded on October 23, 2023, from GitHub (https://github.com/mir-group/nequip) is used for all training and inference. The initial dataset generated as described above was comprised of 1000 structures. This was split into training set (80%) and validation set (20%) and multiple NequIP [44] MLIPs with a 6 Å radial cutoff and equivariant E(3) products up to $L_{max}$ = 2 in the tensor layers were trained to find a MLIP with the optimal set of hyperparameters. The NequIP MLIPs were trained in two stages of 500 epochs each – first where the force to energy ratio in the loss function is 20:10, ensuring the forces are trained accurately, and second where the force to energy ratio is 10:10. In both stages the stress loss coefficient is set high (100,000) to ensure the stresses are also sufficiently learned. Seven of the best MLIPs based on the validation loss we then selected to be used as the committee of MLIPs (See Supporting information for training input files for all committee members). Training and inference of our MLIPs were performed using a NVIDIA RTX 4090 GPU, and 2 AMD EPYC 7551P CPU cores.

For generating new structures to be added to the initial training set, NequIP-based MD simulations were performed using the OpenMM-ML plugin (version 1.1) in the OpenMM[45] code (version 8.1.1) except for contour exploration simulations which were performed in ASE. In addition to this, we also used umbrella sampling to generate specific configurations of ions in bulk water at 0.4 M concentration. All umbrella sampling MD simulations were performed using OpenMM in which the PLUMED[46-48] plugin (version 2.0.1) in OpenMM was used to apply harmonic umbrella potentials of the from $V(r) = k(r_0 - r)^2$ with a force constant of 200 kJ/mol/nm$^2$ on the collective variable (i.e., sodium-sulfate distance, graphene-sodium distance or graphene-sulfate distance). To select new structures that were added to the original training set, we followed a query-by-committee approach whereby critical structures are selected based on uncertainty of the committee of MLIPs in the energy, force predictions (see Supporting Information for more details). In each iteration, we add only 100 to 300 structures to the training and validation sets, respectively. The active learning process is repeated until an MLIP that is stable and sufficiently reproduces bulk properties is obtained. We find an approximately linear correlation between the committee uncertainty and true error (Figure S13), indicating that uncertainty-based selection is suitable for selecting new structures without calculating the ground truth.[19, 49] Our converged MLIPs required 9 iterations with a final training set of 4,047 structures and a validation set of 2,213 structures while covering all relevant distances and bond lengths in this system (See Figures S14 to S17). Finally, to generate an independent test set, we ran MD simulations for each subsystem. For bulk water and the bulk electrolyte solution, we ran 500 ps NPT simulations with our final MLIP using 128 water molecules and 0 to 5 $Na_2SO_4$ (0 to 2 M concentration) using the same setup described above. For the slab models, we ran 500 ps NVT simulations using 128 molecules and 0 to 5 $Na_2SO_4$ (0 to 2 M concentration) with different initial configurations. In addition to this, we also ran NVT simulations whereby water or the solution is confined between two graphene sheets to evaluate the performance of our model at the interfaces. To ensure that selected structures are uncorrelated, we selected every 400$^{th}$ structure from each MD and ended up with a total of 42,134 structures. These structures were then labeled at DFT level and used as an independent test set to evaluate the accuracy of our MLIP. The final MLIP energy MAE is trained to 0.53 meV/atom and the force MAE is trained to 29 meV/Å in the total test set (See Supporting Information for correlation and distribution plots of energy and force errors for each subset of test data).

3.3. MLIP-based MD

To illustrate our iCµMD simulation approach using MLIPs, we use the aqueous Na$_2$SO$_4$-graphene interface as a model. We selected this system because of the added complexity of modeling the polyatomic divalent sulfate ion which has been shown in previous studies to significantly alter the structure of the graphene-water interface.[50] For production runs, we carried out MD simulations using OpenMM in simulation cells containing 1,600 water molecules between two fixed graphene sheets whereby the volume of the solution was chosen to match a water density of 1 g/cm$^3$. The simulation cell had dimensions of 34.15 Å x 29.57 Å x 110.00 Å and for the first iteration contained a total of 5,785 atoms comprising of 2 graphene sheets with 384 carbons each (768 carbon atoms in total), 1,600 water molecules (4800 atoms in total), 31 SO$_4^{2-}$ ions (155 atoms in total), and 62 Na$^+$ ions (Figure 2). Five replicates of the initial solution were created using Packmol[24] at a nominal concentration of 1 M. After packing the simulation box using Packmol, the system was geometry optimized and equilibrated within the NVT ensemble under Langevin dynamics and a 1 fs timestep at 310 K for 300 ps with the hydrogen mass set to 4 amu. After this initial equilibration, the hydrogen mass reset to 1 amu and NVT simulations were run with a 0.5 fs timestep at 300 K for at least 4 ns to ensure efficient sampling of the bulk and interfacial regions. Here, the bulk region is defined as ±7 Å from the center of the solution (i.e., a 14 Å region) and we measure the bulk concentration every 500 ps. We use a slightly larger bulk region in this example to smooth out fluctuations due to the larger size of the sulfate ion. It should be noted that we take the position of the sulfur atom as the position of the sulfate ion for the purpose of analyzing the sulfate ion distribution. It should also be noted that the ion density in the Z-direction is taken in 1 Å bins. Error bars on the Z-direction concentration profile are calculated as the standard error of the mean (SEM) as a function of Z position using the following equation:

$$SEM(z) = \frac{\sigma(z)}{\sqrt{N}} \qquad (2)$$

where σ(z) is the standard deviation at a given position and N is the number of simulations.

To define the iteration stopping criteria, we use a moving average window of 5 ns, and a slope threshold of ±2 x 10$^{-3}$ M/ns for the Na$_2$SO$_{4,aq}$-graphene. Additionally, we require that the slope remains within the threshold for a duration of at least 15 ns to avoid transient convergence. Finally, to define the stopping criteria for the iCµMD simulations, we stop iterating when the bulk concentration is within ±0.03 M of the target concentration based on the uncertainty defined previously by Finney et al.[10]

4. Results and Discussion

In this section, we will first discuss the classical force field-based iCµMD results followed by the MLIP-based iCµMD results. We begin by discussing the time evolution of bulk concentration in the NaCl$_{aq}$-graphite and NaCl$_{aq}$ - air systems across the iCµMD iterations. We then briefly discuss the ion distribution including the interfacial and bulk regions. Next, we will discuss the MLIP-based iCµMD and analyze the time evolution of bulk concentration in the Na$_2$SO$_{4,aq}$ – graphene system as well as the concentration of ion in the interfacial and bulk regions. We also highlight the advantages and efficacy of the iCµMD simulation which converges within only two iterations.

## 4.1. Classical force field-based iCμMD

Figure 3 (left) shows the variation of the bulk concentration with time in the $NaCl_{aq}$- graphite system across two iterations. For the first iteration, we observe that equilibration is reached in about 30 ns as indicated by the slope of the bulk concentration in Figure S18 (left) whereby the bulk concentration dropped to 0.85 M from the nominal 1 M concentration (220 ion pairs) due to the accumulation of ions at the interface. Using equation 1, we then calculated the required number of ions to reach the target concentration and added 39 more ion pairs following the procedure outlined in the computational methods. We then initialize the next iteration using an already well-equilibrated structure from the initial iteration and as shown in Figure S18 (right), the bulk concentration equilibrates after about 53 ns of the second iteration resulting in a bulk concentration of 0.99 M which remains stable (within ±0.03 M of the targeted 1 M bulk concentration) even after 300 ns of simulation. Therefore, based on our iCμMD stopping criteria we consider the $NaCl_{aq}$-graphite system converged after only two iterations and ~ 80 ns of simulation. Additionally, Figure 3 (right) and Figure S19 also shows the Z-direction distribution of ions in the last 200 ns of the converged second iteration while the concentration of ions at the respective interfaces increased proportionally compared to the first iteration. The $Na^+$ concentration at the negative electrode increased from 11.26 M to 14.30 M, while the $Cl^-$ concentration at the positive electrode increased from 15.74 M to 18.58 M.

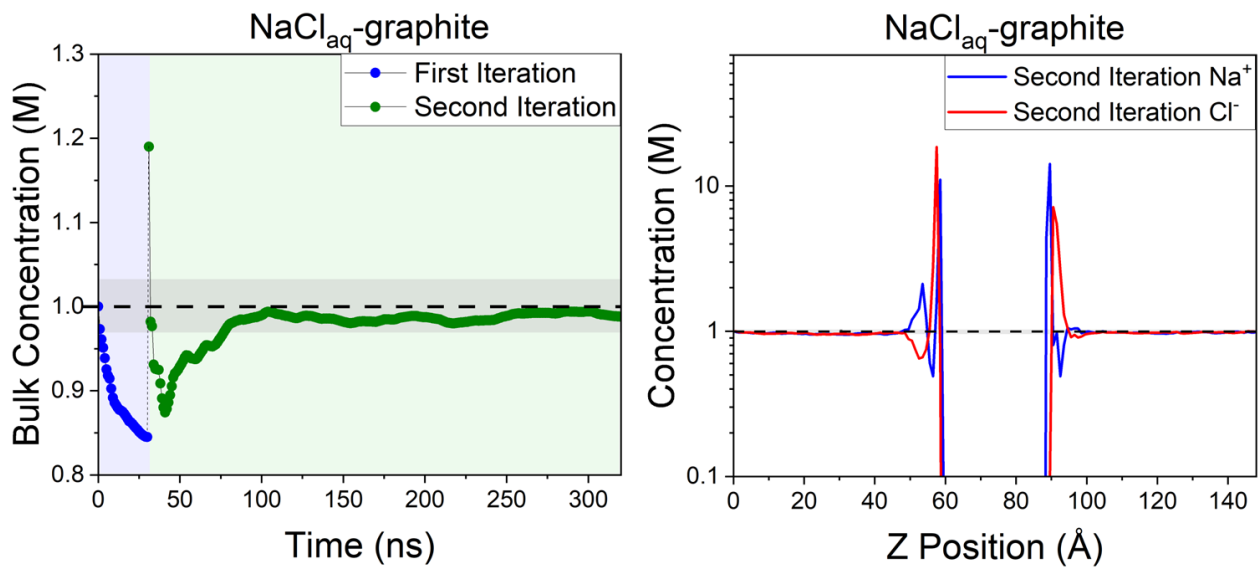

**Figure 3**. The moving average of the molar concentration of ions in the $NaCl_{aq}$- graphite system as a function of simulation time within the bulk region (left), and molar concentration of ions in the $NaCl_{aq}$-graphite system after 300 ns as a function of Z position in the simulation cell in the second iteration (right). Note that a moving window of 5 ns is used to calculate the moving average of the bulk concentration. Also note that the y-axis on the right plot is on a log scale. The bottom and top graphene layers are positioned at 62 Å and 86 Å in the simulation cell, respectively. For visualization purposes, the ion distribution is an average over the last 200 ns of the simulation. The horizontal dashed black line denotes the target 1 M bulk concentration while the grey shaded region represents the ±0.03 M window for the iCμMD convergence.

Figure 4 (left) shows the variation of the bulk concentration with time in the $NaCl_{aq}$ - air system across two iterations. For the first iteration, we observe that equilibration is reached in about 23 ns as indicated by the slope of the bulk concentration in Figure S20 (left) whereby the bulk concentration increased to 1.07 M from the nominal 1 M concentration (220 ion pairs) due to the movement of the ions away from the air interface and into the bulk region. Using equation 1, we then calculated the required number of ions to reach the target concentration and removed 17 ion pairs following the procedure outlined above. We then initialize the next iteration using the already well-equilibrated structure from the initial iteration and as shown in Figure S20 (right), the bulk concentration equilibrates after about 83 ns of the second iteration resulting in a bulk concentration of 1.02 M which remains stable (within ±0.03 M of the targeted 1 M bulk concentration) even after 300 ns of simulation. Therefore, based on our iCμMD stopping criteria we consider the $NaCl_{aq}$ - air system converged after only two iterations and ~ 100 ns of simulation. Figure 3 (right) and Figure S21 also shows the Z-direction distribution of ions in the last 200 ns of the converged second iteration and here we find that the $Na^+$ ions have a greater affinity for the air interface compared to $Cl^-$ ions.

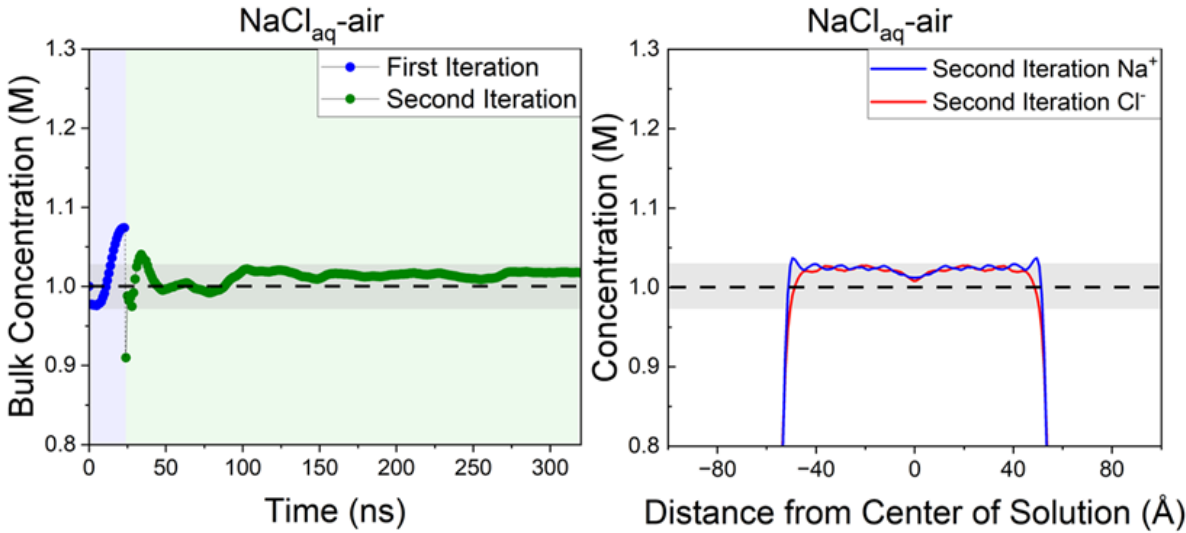

**Figure 4**. The molar concentration of ions in the $NaCl_{aq}$ - air system as a function of simulation time within the bulk region (left), molar concentration of ions in the $NaCl_{aq}$ - air system after 300 ns as a function of Z position in the simulation cell in the second iteration (right). Note that a moving window of 5 ns is used to calculate the moving average of the bulk concentration. For visualization purposes, the ion distribution is an average over the last 200 ns of the simulation. The horizontal dashed black line denotes the target 1 M bulk concentration while the grey shaded region represents the ±0.03 M window for the iCμMD convergence.

4.2. iCμMD results based on MLIPs

Figure 5 (left) shows the variation of the bulk concentration with time in the $Na_2SO_{4,aq}$-graphene system across two iterations. For the first iteration, we observe that equilibration is reached in about 20 ns based on our convergence criteria and as seen in Figure S22 (left) whereby the bulk concentration initially increased to 1.2 M from the nominal 1 M concentration before slowly equilibrating to 1.07 M as ions accumulate at the interfaces. Using equation 1, we then calculated the required number of ions to reach the target concentration and removed 2 ion triplets following the procedure outlined above. We then initialize five replicates of the next iteration using the already well-equilibrated structure from the initial iteration and as shown in Figure S22 (right), the bulk concentration quickly equilibrates after about 11 ns resulting in a bulk concentration of 0.98 M which remains stable (within ±0.03 M of the targeted 1 M bulk concentration) even after 50 ns of simulation. Therefore, based on our iCμMD stopping criteria we consider the $Na_2SO_{4,aq}$ - graphene system converged after only two iterations and ~ 70 ns of simulation. Figure 5 (right) also shows the Z-direction distribution of ions in the last 50 ns converged second iteration and here we also find that the $Na^+$ ions have a greater affinity for the graphene interface compared to $SO_4^{2-}$ ions. Additionally, while there is no charge on the graphene sheets, we observe that $SO_4^{2-}$ preferentially accumulates near the interface rather than in the bulk given that the concentration of $SO_4^{2-}$ ions near the interface is approximately twice the bulk concentration. Comparatively, the $Na^+$ surface concentration is only 1.3 times the bulk concentration. Therefore, this may suggest that $SO_4^{2-}$ ions have a strong interaction with the graphene but their large and ordered water coordination shell[51-54] precludes their close proximity to the graphene sheet.

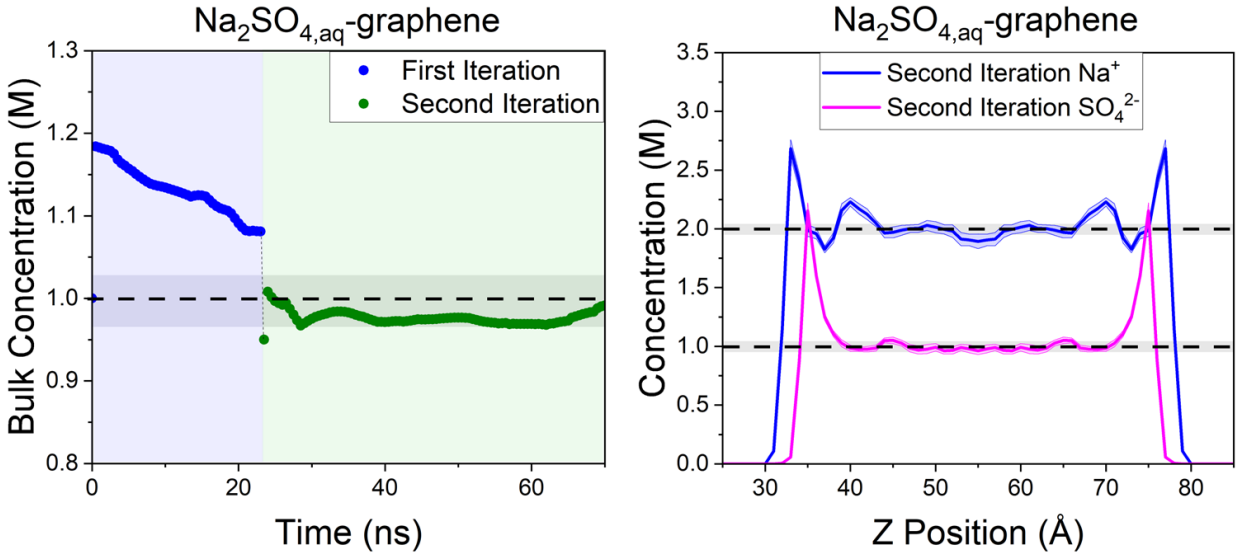

**Figure 5**. The molar concentration of ions in the $Na_2SO_{4,aq}$ - graphene system as a function of simulation time within the bulk region (left), symmetrized molar concentration of ions in the $Na_2SO_{4,aq}$ - graphene system after 50 ns as a function of Z position in the simulation cell in the second iteration (right). The bottom and top graphene layers are positioned at 28.25 Å and 81.75 Å, respectively. Note that a moving window of 5 ns is used to calculate the moving average of the bulk concentration. The horizontal dashed black line denotes the target 1 M bulk concentration (2 M for $Na^+$ ions) while the grey shaded region represents the ±0.03 M window for the iCµMD convergence. The pink and blue shaded regions in the right plot represent the error bars of the concentration of $SO_4^{2-}$ and $Na^+$ ions, respectively based on the average concentration over 5 independent MD simulations.

4. Conclusions

In this work, we have introduced the iCµMD simulation method, a simple yet practical simulation approach for satisfying a constant chemical potential condition in the simulation of interfacial systems. By iterating a 'run-measure-update' loop until the bulk concentration converges to the desired target concentration, we demonstrated that iCµMD can achieve a constant-µ condition in a few iterations using a straightforward MD setup. Compared to the CµMD approach, iCµMD offers clear advantages due to its ease of setup, no need for the interfacing multiple of codes, and direct compatibility with both classical force fields and MLIPs. To demonstrate our method, we modeled the $NaCl_{aq}$-graphite and $NaCl_{aq}$-air interfaces using classical force fields, and the $Na_2SO_{4,aq}$-graphene interface using MLIPs and set a target bulk concentration of 1 M. Our results showed that for each system, only two iterations were required to reach the target concentration. Additionally, a total of 50 ns, 100 ns, and 70 ns were required to reach the target concentration (within ± 0.03M) in the the $NaCl_{aq}$-graphite, $NaCl_{aq}$-air, and $Na_2SO_{4,aq}$-graphene systems, respectively. Importantly, we also demonstrated that the iCµMD method can be integrated with MLIP-driven simulations which enable DFT-level accuracy constant potential simulations.

We have shown that the iCµMD framework extends to solid-liquid and liquid-air, however, it is also applicable to solid-gas, liquid-liquid systems, and mixed-solute systems. Its generality and MLIP compatibility make iCµMD an attractive tool for simulating any fluid interface where control of chemical potential is required, while its flexibility allows the user to set convergence criteria and thresholds to suit their needs. iCµMD can be readily integrated into existing MD workflows and even specialized simulations such as enhanced sampling simulations. By providing a simple method of performing constant-µ simulations across a wide range of systems, we anticipate adoption of iCµMD in studies of interfacial chemistry, materials science, and beyond as its only requirement is to be able to converge interfacial simulations with a bulk region.

**Supplementary Material**

The supplementary material includes figures on the comparison MLIP and DFT energy and forces in the test set, distribution plots showing distances and bond lengths captured within the training and test sets, and figures showing the convergence of bulk concentration and concentration profiles across the iCµMD iterations.

**Acknowledgements**

This work was made possible by the U.S. Department of Energy, Office of Science, under Award DE-SC0024654. The authors thank Tristan Maxson, Sophia Ezendu, Gbolagade Olajide, and Mustapha Iddrisu for their insightful comments on the manuscript and work. A.S would like to acknowledge the financial support of the University of Alabama Graduate School as a Graduate Council Fellow. This work was also made possible in part by a grant of high-performance computing resources and technical support from the Alabama Supercomputer Authority. This research used resources of the National Energy Research Scientific Computing Center (NERSC), a U.S. Department of Energy Office of Science User Facility located at Lawrence Berkeley National Laboratory, operated under Contract DE-AC02-05CH11231 using NERSC Award BES-ERCAP0024218. Any opinions, findings, conclusions, and/or recommendations expressed in this material are those of the authors(s) and do not necessarily reflect the views of funding agencies.

**Author Declarations**

Conflict of Interest

The authors have no conflicts of interest to disclose

Author Contributions

Ademola Soyemi: Investigation (lead); Methodology (equal); Formal analysis (equal); Writing – original draft (lead); Writing – review and editing (equal). Khagendra Baral: Investigation (supporting); Formal analysis (supporting); Tibor Szilvási: Conceptualization (lead), funding acquisition (lead); Writing – review and editing (equal); Methodology (equal), Formal analysis (equal), Supervision (lead).

**Data Availability**

The data that support the findings of this study are available from the corresponding author upon reasonable request. In addition, the full training, validation, and test sets and input files for training the NequIP models can be downloaded from our Zenodo repository (DOI: https://doi.org/10.5281/zenodo.15512423). We also provide python pseudocode which outlines the iCµMD procedure in the Zenodo repository.